\documentclass[a4paper,12pt]{spieman}  
\usepackage{amsmath,amsfonts,amssymb}
\usepackage{graphicx}
\usepackage{setspace}
\usepackage{tocloft}
\usepackage[colorlinks=true, allcolors=blue]{hyperref}
\usepackage{threeparttable,subcaption}
\usepackage{comment}
\usepackage{here}
\usepackage{siunitx}
\usepackage{multirow}
\usepackage{bm}

\title{Ground test results of the electromagnetic interference for the x-ray microcalorimeter onboard XRISM}
\author[a,b,*]{Miki Kurihara}
\author[b]{Masahiro Tsujimoto}
\author[c]{Megan E. Eckart}
\author[d]{Caroline A. Kilbourne}
\author[b]{Frederick T. Matsuda}
\author[d]{Brian McLaughlin}
\author[b]{Shugo Oguri}
\author[d]{Frederick S. Porter}
\author[e]{Yoh Takei}
\author[ ]{on behalf of of the XRISM \textit{Resolve} team}
\author[f]{Yoichi Kochibe}

\affil[a]{Graduate School of Science, The University of Tokyo, Bunkyo-ku, Tokyo, 113-0033 Japan}
\affil[b]{Institute of Space and Astronautical Science, Japan Aerospace Exploration
Agency, Chuo-ku, Sagamihara, Kanagawa, 252-5210 Japan}
\affil[c]{Lawrence Livermore National Laboratory, Livermore, CA 94550, USA}
\affil[d]{National Aeronautics and Space Administration (NASA), Goddard Space Flight Center, Greenbelt, MD 20771, USA}
\affil[e]{Institute of Space and Astronautical Science, JAXA, Tsukuba, Ibaraki 305-8505, Japan}
\affil[f]{Fujitsu Limited, Saiwai-ku, Kawasaki, Kanagawa, 212-0014 Japan}

\cftpagenumbersoff{figure}
\cftpagenumbersoff{table} 
\begin{document} 
\maketitle

\begin{abstract}
 Electromagnetic interference (EMI) for low-temperature detectors is a serious concern
 in many missions. We investigate the EMI caused by the spacecraft components to the
 x-ray microcalorimeter of the \textit{Resolve} instrument onboard the X-Ray Imaging and
 Spectroscopy Mission (XRISM), which is currently under development by an international
 collaboration and is planned to be launched in 2023. We focus on the EMI from (a) the
 low-frequency magnetic field generated by the magnetic torquers (MTQ) used for the
 spacecraft attitude control and (b) the radio-frequency (RF) electromagnetic field
 generated by the S and X band antennas used for communication between the spacecraft
 and the ground stations.  We executed a series of ground tests both at the instrument
 and spacecraft levels using the flight-model hardware in 2021--2022 in a JAXA facility
 in Tsukuba. We also conducted electromagnetic simulations partially using the Fugaku
 high-performance computing facility. The MTQs were found to couple with the
 microcalorimeter, which we speculate through pick-ups of low-frequency magnetic field
 and further capacitive coupling. There is no evidence that the resultant energy
 resolution degradation is beyond the current allocation of noise budget. The RF
 communication system was found to leave no significant effect. We present the result of
 the tests and simulation in this article.
\end{abstract}


{\noindent \footnotesize\textbf{*}Miki Kurihara,  \linkable{kurihara@ac.jaxa.jp} }

\begin{spacing}{1}   

\section{Introduction}\label{s1}
Electromagnetic interference (EMI) is a growing concern in modern astronomical
instruments with increasing demands for high sensitivity and low noise. One of the areas
of serious concern is the low-temperature detectors in space-borne or air-borne
platforms. A noise level of $\mathcal{O}(10^{-18}~\textrm{W}/\sqrt{\mathrm{Hz}})$ is
required in a densely packed test bed consuming
$\mathcal{O}(10^{3}~\mathrm{W})$. Examples can be found in Planck\cite{tauber2010}
high-frequency instrument\cite{lamarre2010} and SPIDER\cite{gualtieri2018} for cosmic
microwave background observations and ASTRO-H\cite{takahashi2016} soft x-ray
spectrometer (SXS)\cite{kelley2016,mitsuda2014} for x-ray observations. More examples
will follow in future missions.

\underline{\textit{Resolve}}\cite{ishisaki2022} x-ray microcalorimeter for the X-Ray Imaging and
Spectroscopy Mission (XRISM)\cite{tashiro2020} was designed to be almost the same as the
SXS\cite{kelley2016,mitsuda2014} for ASTRO-H\cite{takahashi2016} to recover its
excellent science programs yet to be achieved due to the unexpected early loss of the
mission by the malfunction of the spacecraft attitude control in March 2016. The SXS
suffered from the EMI from the bus system that degraded the detector
performance\cite{eckart2018}, which was only recognized after the integration of the
instrument into the spacecraft. Still, the SXS demonstrated its excellent performance
beyond the requirement in the orbit\cite{porter2018,leutenegger2018}.

Based on the lessons of the SXS, we started with an EMI verification program for
\textit{Resolve} and conducted a series of tests in 2021--2022 at JAXA's Tsukuba space
center both at the instrument and spacecraft levels. We recently finished the major part
of these tests in August 2022. The purpose of this article is to report the results of
the EMI ground tests and simulations so that it will be a reference in interpreting the
in-orbit data with \textit{Resolve} and in designing future instruments susceptible to
EMI. Amongst various EMI effects, we focus on (i) the radiative EMI by the low-frequency
magnetic field and (ii) the high-frequency electromagnetic field. The former was
observed in the SXS\cite{eckart2018}, while the latter remains unverified in the
SXS. The conductive EMI was also tested, which is described in a separate
article\cite{ishisaki2022}.

The article is structured as follows. In \S~\ref{s2}, we give an overview of the
\textit{Resolve} instrument (\S~\ref{s2-1}) and the spacecraft (\S~\ref{s2-2}). The
description is focused on the victims of the EMI, which are the microcalorimeter and
anti-coincidence detectors and their signal chain (\S~\ref{s2-1}) and on the
perpetrators of the EMI, which are the spacecraft attitude control and communication
systems (\S~\ref{s2-2}). In \S~\ref{s3}, we present the low-frequency magnetic field EMI
results of the simulations (\S~\ref{s3-1}), instrument-level test (\S~\ref{s3-2}),
spacecraft-level test (\S~\ref{s3-3}), and discuss the coupling mechanism
(\S~\ref{s3-4}). In \S~\ref{s4}, we present the high-frequency electromagnetic field EMI
result in the same structure: simulation (\S~\ref{s4-1}), instrument-level test
(\S~\ref{s4-2}), spacecraft-level test (\S~\ref{s4-3}), and discussion about the outcome
(\S~\ref{s4-4}). The article is summarized in \S~\ref{s5}.

\section{\textit{Resolve} onboard XRISM}\label{s2}
\subsection{\textit{Resolve}}\label{s2-1}
\textit{Resolve}\cite{ishisaki2022} is one of the two science instruments
onboard XRISM\cite{tashiro2020}, which aims to achieve an energy resolution of 7~eV
(FWHM) at 5.9~keV non-dispersively over a wide range of energy
(0.3--12~keV). \textit{Resolve} hosts an array of 6$\times$6 x-ray microcalorimeter
pixels thermally anchored to the 50~mK heat bath with a thermal time constant of
3.5~ms\cite{kilbourne2018a}. In each pixel, a HgTe absorber absorbs incoming x-ray photons and the resultant
temperature rise is measured by a Si thermistor with a temperature-dependent impedance
of $\sim$30 M$\Omega$. Each microcalorimeter pixel is biased in series with a 140 M$\Omega$
load resistor, and the change in the voltage across the thermistor is the signal. Below
the microcalorimeter array, an anti-coincidence
detector\cite{kilbourne2018a} (anti-co) is placed for identifying particle events.
The anti-co is a Si ionization detector biased in series with a 2.5 M$\Omega$ load
resistor; the voltage across the load resistor is the signal, which is zero in
quiescence. The bias voltage for the microcalorimeter pixels, but not the anti-co, is divided
down by a factor of 121 between the connector at the dewar main shell and the top of
load resistors\underline{\cite{Chiao2018}}. The nine pixels within each quadrant of the array are connected to the
same bias line. Each of these high impedance signal is converted into low impedance by a
junction field effect transistor (JFET) source follower. The JFETs must be operated at
130 K, and, thus, are thermally isolated from the cold stage. Because of the
commonalities and differences of the microcalorimeter and anti-co detectors, observation
of the differential response is useful for evaluating potential coupling mechanisms of
the noise.

The JFET signals are passed to \underline{the x-ray amplifier box (XBOX)}\cite{kelley2016}, which AC-couples, amplifies,
and applies an anti-aliasing filter before digitizing them at 12.5 Hz sampling. The
digitized signal is relayed to other room-temperature electronics called \underline{the
pulse shape processor (PSP)}\cite{ishisaki2018}, which is responsible for x-ray event detection and
reconstruction as well as collecting the detector noise data. We use the spectra made
from noise records of an 8k sample length (0.65536 s) for the frequency-domain data and
dumps of continuous 50k samples (4.096~s) synchronous among all
microcalorimeter and the anti-co channels for the time-domain data to evaluate the
detector responses to EMI.

The detectors and JFETs are housed inside the dewar\cite{fujimoto2017,yoshida2018}.  A
tank with superfluid He provides a stable thermal anchor of $\sim$1.2~K.  From there,
two adiabatic demagnetization refrigerators (ADRs) work in series to cool the detector
stage to 50~mK\cite{shirron2016}, controlled by room-temperature electronics called the
ADR controller (ADRC)\cite{kelley2016}. Between the detector stage and the dewar main
shell are multiple isolated thermal shields made of Al of a few mm thickness. The dewar
interior is cooled actively by five cryocoolers\cite{sato2012} and passively by the He
vapor cooling, and its exterior is passively cooled by radiative cooling toward deep
space. In case of the depletion of superfluid He, a third ADR works to cool the He tank
for an extended lifetime\cite{sneiderman2018,kanao2017}.

The dewar is an Al vacuum vessel, leak-tight on the ground under air, thus constituting
a Faraday cage against the external EMI environment. For x-ray observations in orbit,
the dewar needs to be open along the x-ray light path. An apparatus called the gate
valve (GV) is installed at the top of the dewar, which is kept closed on the ground and
during launch. The GV will be permanently opened using a non-explosive actuator during
the commissioning phase after the spacecraft out-gassing settles. The GV door has a Be
window of $\sim$270~$\mu$m thickness\cite{Midooka2021} to transmit x-rays above
$\sim$2~keV.

\subsection{Spacecraft}\label{s2-2}
\begin{figure}[htbp!]
 \centering
 \includegraphics[width=0.9\textwidth]{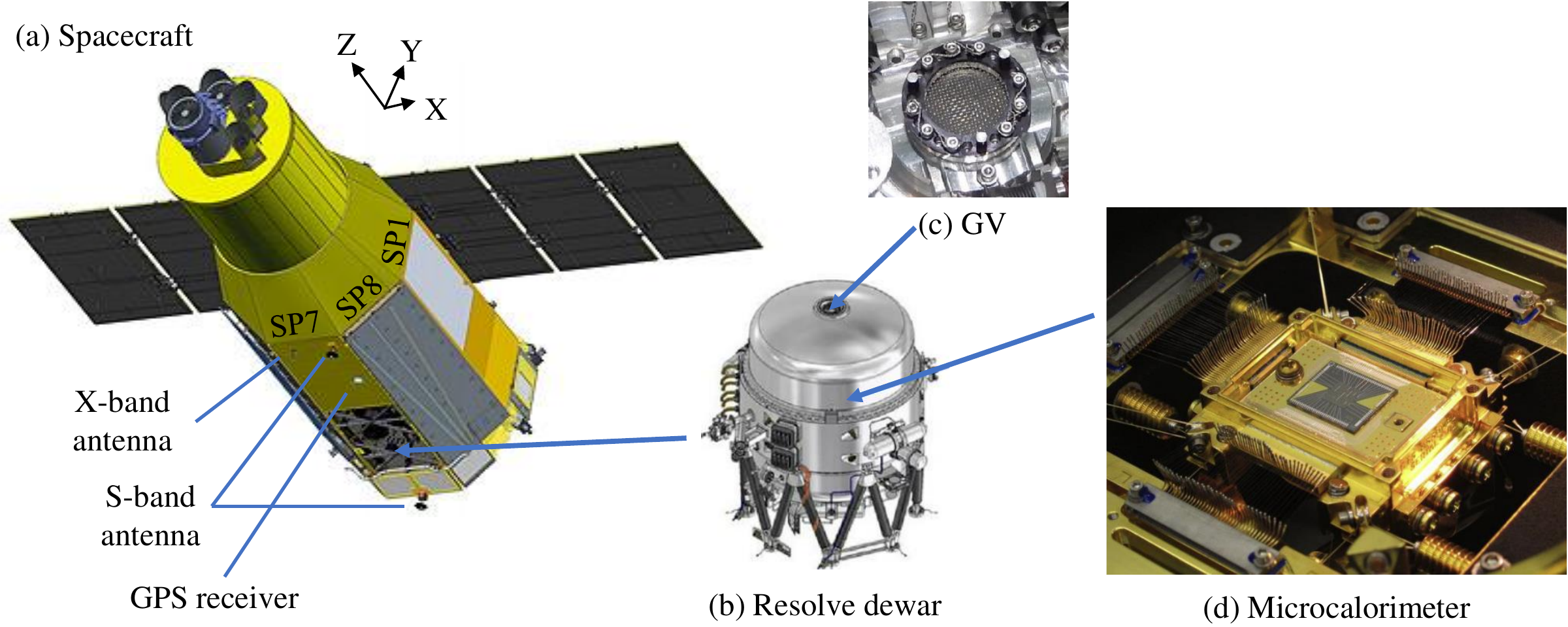}
 \caption{Illustration of the spacecraft and the \textit{Resolve} instrument. (Photo (c) is from Ishisaki et al.\cite{ishisaki2022})}
 \label{f-spacecraft}
\end{figure}

XRISM is planned to be launched in 2023 from the JAXA's Tanegashima space center into a
near-Earth orbit with an altitude of 575~km and an inclination of 31 degrees. The
spacecraft is in the final integration testing, as of writing, since April 2022 at the
JAXA's Tsukuba space center. The structure of the spacecraft inherits the design of
ASTRO-H (Figure \ref{f-spacecraft}). It has a weight of $2.3 \times 10^{3}$~kg and an
envelope size of 7.9, 9.2, and 3.1~m respectively for the height ($z$), length ($x$),
and width ($y$). The main body is composed of eight side panels (SP1--8) of
990$\times$3100~mm$^{2}$ in size, whose inside has the top, middle, and lower plates
perpendicular to the SPs standing on the bottom structure. For \textit{Resolve}, the x-ray
mirror is placed on the top panel, the dewar is upright on the bottom structure, and
room-temperature electronics are inside the body on the SPs. The solar array panels are
stored in SP2 and SP4 at launch and are deployed in parallel with SP3, which are oriented
toward the Sun. A part of SP7 is a lattice window exposed to deep space to cool the
dewar surface radiatively.

For the attitude control of the spacecraft, the reaction wheels (RWs) and the magnetic
torquers (MTQs) are used as main actuators. The RWs are composed of four units with 
the base momentum of a 3000~rpm rotation, which are used to rebalance the angular momentum 
within the spacecraft for pointed observations and maneuvering. The MTQs dump the accumulated
angular momentum against the Earth magnetic field.
Three MTQs are installed inside SP3 for the $x$ axis and SP5 for the $y$ and $z$
axes. Each MTQ is a solenoid of a 920~mm length and a 34~mm radius, which generates the
maximum magnetic moment of 900~A~m$^{2}$ for a $\pm$35~V bipolar DC drive. The field
strength is controlled by changing the duty ratio of the DC with pulse width modulation
(PWM) at 127~Hz frequency. This is close to the thermal time constant of the detector,
and can cause significant degradation when coupled. This was indeed observed in the
SXS\cite{eckart2018}, but its coupling path (magnetic or conductive) has not been
clarified.

For communication with the ground, the spacecraft has four S-band and two X-band
antennas. A half of them (two S-band and one X-band antennas) are located outside of
SP7, while the other half are outside of SP3. The S-band is used both for uplink and
downlink at 2~GHz, while the X-band is used only for downlink at 8~GHz. All of them use
a cross-dipole antenna with a reflector to increase the forward-to-backward ratio. Another
use of radio frequency (RF) is the GPS receivers outside of SP3 and 7. The electronics
for the RF modulation/demodulation, filtering, and amplifying are installed inside of
the SPs. These RF equipments, in particular those for downlink with strong emission
local to the spacecraft, are the sources of high-frequency electromagnetic
field. Because the \textit{Resolve} dewar constitutes a Faraday cage with the GV closed
on the ground and early in the orbit, the RF signals would not affect the detector
inside the dewar. This was indeed the case for the SXS, which ended its life before
the GV was opened in the orbit. However, when the GV will be opened for
\textit{Resolve}, the cage is broken with a $\sim 35$~mm diameter hole left open and we may
expect RF interference with a frequency higher than its cut-off at $\sim$2~GHz. A
particular concern is the RF emission from the X-band or S-band antenna outside of SP7,
which can diffract into the spacecraft main body through the opening in SP7 and reach
the detector through the opened GV. The modulation in the carrier frequency used for
communication may load energy to the detector that varies in time within its
bandpass. This could not be verified for the SXS and thus must be investigated
for \textit{Resolve}.

\section{Magnetic EMI}\label{s3}

\subsection{Simulation}\label{s3-1}
\begin{figure}[htbp!]
 \centering
 \includegraphics[width=0.9\textwidth,clip]{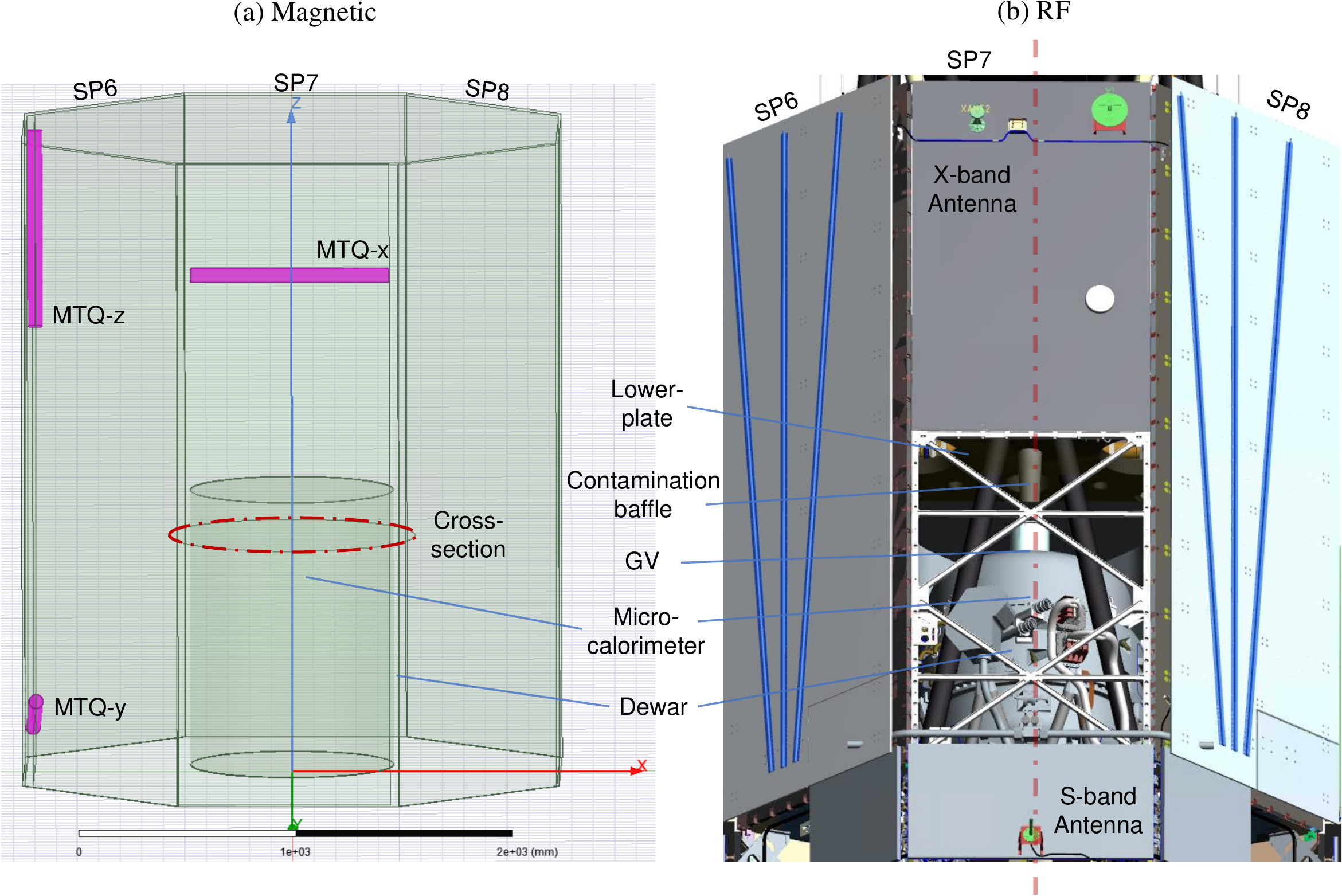}
 \caption{Simulation models of (a) magnetic and (b) RF EMI seen from the SP7 in
 a slightly upward direction. The emission sources (a; three MTQs and b; two RF
 antennas) and the evaluation plane of the simulation (a; circular cross section
 perpendicular to the $z$ axis including the feed-through and b; cross section
 perpendicular to the $x$ axis including the detector center) are shown in red
 dashed-and-dotted curve or line.}
 \label{f-simulation}
\end{figure}

We start with the simulation of the magnetic field generated by the three MTQs. The MTQ
has a large inductance (6.7~H) with a cut-off frequency of 0.68~Hz, which is much
smaller than its PWM drive frequency. The magnetic field is approximated as DC, thus we
calculated the static solution of Maxwell's equations. The AC component can be
calculated by scaling the DC simulation results in the post-processing. We used the
\texttt{Maxwell} software\footnote{See
\url{https://www.ansys.com/products/electronics/ansys-maxwell} for detail.} provided by
ANSYS based on the finite element method (FEM) solver. A model was made by simplifying
the spacecraft main body and the \textit{Resolve} dewar (figure~\ref{f-simulation} a). A
personal computer is sufficient for this simulation.

\begin{figure}[htbp!]
 \centering
 \includegraphics[width=0.9\columnwidth,clip]{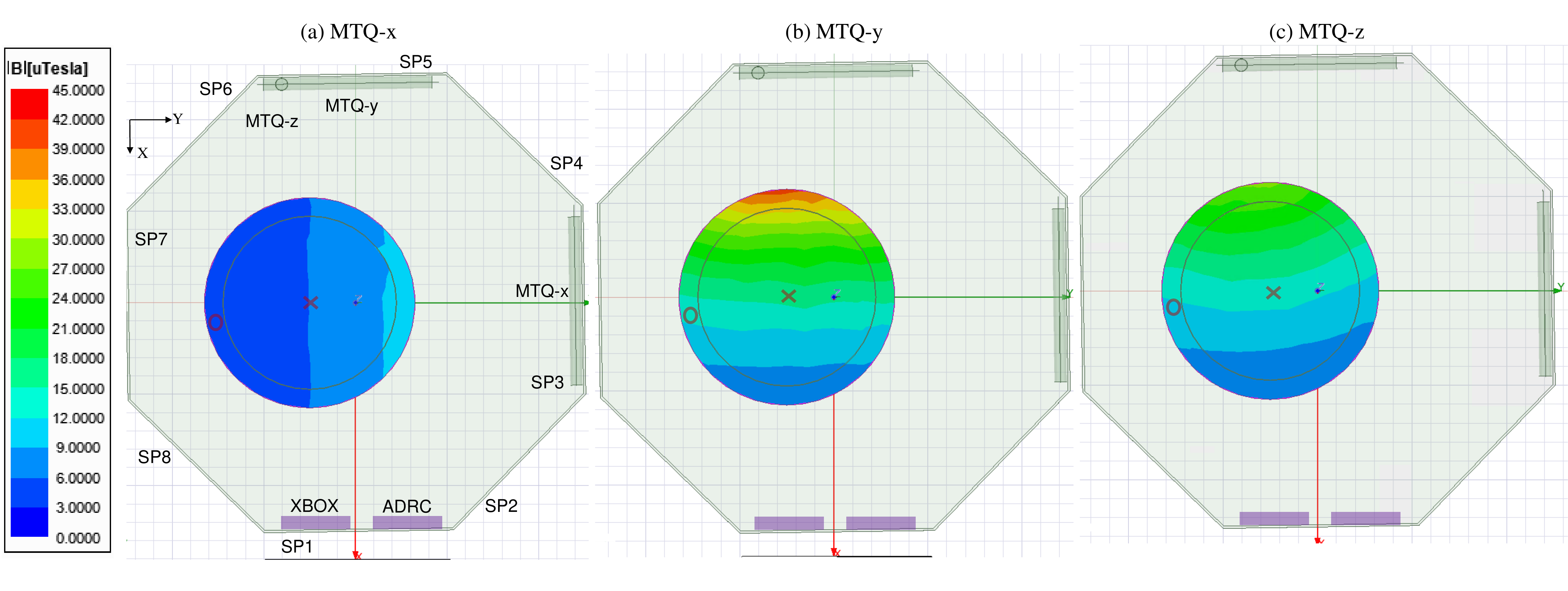}
 \caption{Results of the magnetic simulation. The magnetic field strength $\bm{|B|}$ is
 given in the unit of $\mu$T on the plane perpendicular to the $z$ axis at the height of 
 the feed-through of the dewar harness leading to the ADRC/XBOX (Figure~\ref{f-simulation} a). The
 position of the detector and the feed-through is shown with the cross and the circle,
 respectively.}
 \label{f-mag-simresult}
\end{figure}

The results are shown in Figure \ref{f-mag-simresult} for the three MTQs separately,
which generates the magnetic field of different strengths of $\mathcal{O}$(10~$\mu$T) and
orientations around the dewar. A conceivable part to pick up the magnetic field is the 
harness from the room-temperature
electronics (XBOX and ADRC) that goes into the cold stages inside the dewar. The
cross-section of the dewar including their feed-through is chosen for visualizing the
calculated field.

\subsection{Instrument-level test}\label{s3-2}
We performed the magnetic EMI test during the instrument-level test using the
flight-model hardware of \textit{Resolve} on September 14--15, 2021. The test was
designed to measure the detector response against the magnetic field injection. Despite
the nearly DC behavior of the magnetic field, its AC component is more important in the
microcalorimeter bandpass. We need to simulate the PWM shape precisely in the time
domain. We thus used the engineering model (EM) unit of MTQ left from ASTRO-H, which is
the same design as the ones used for XRISM, and drove it in the same way as the flight
MTQ driver does.

\begin{figure}[htbp!]
 \centering
 \includegraphics[width=0.9\columnwidth,clip]{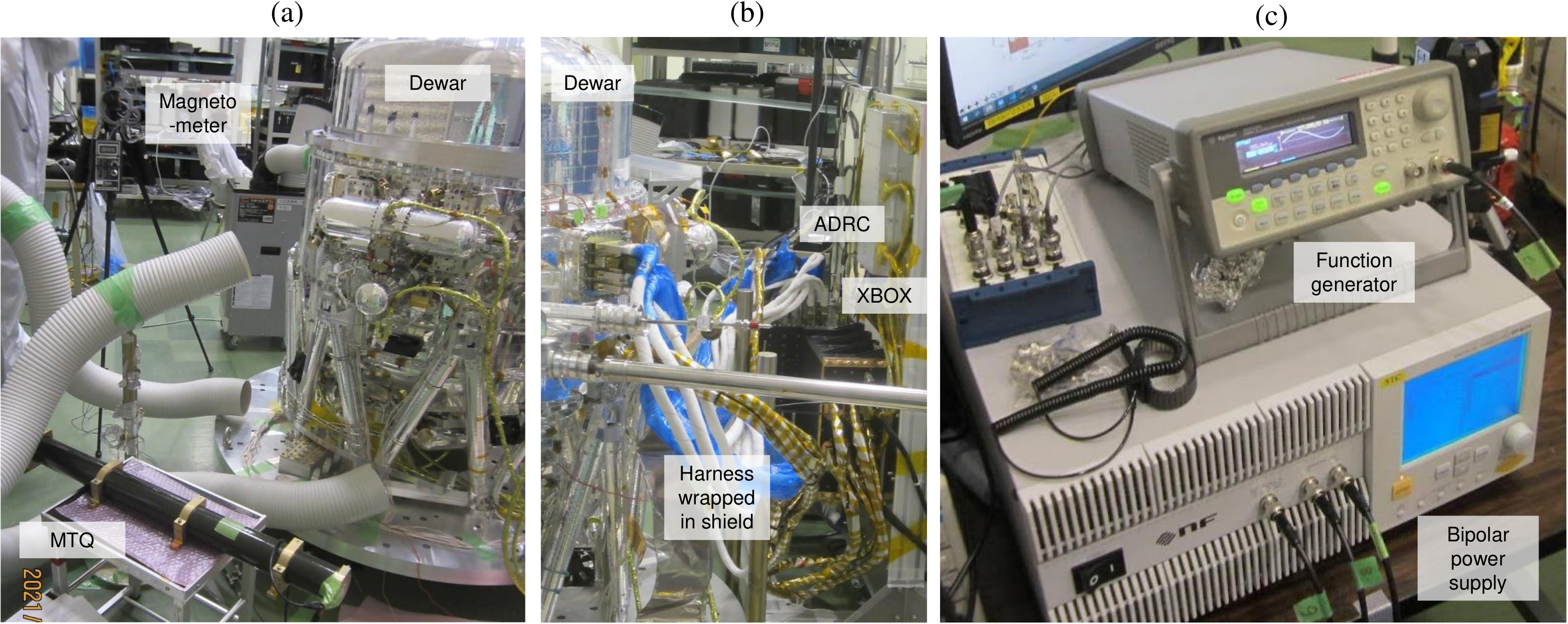}
 \caption{Setup for the magnetic EMI instrument-level test: (a) SP5 and (b) SP8 side of
 the dewar (Figure~\ref{f-mag-simresult}), and (c) drive electronics.}
 \label{f-mag_setup}
\end{figure}

Figure \ref{f-mag_setup} shows the setup of the instrument-level test for the magnetic
EMI. The EM MTQ was placed at a flight position of the MTQ-$y$ relative to the dewar
(Figure \ref{f-mag_setup} a). The PWM shape was made with a function generator, which was
amplified by a bipolar power supply (Figure \ref{f-mag_setup} c). The magnetic field was
measured using magnetometers sensitive to the AC field up to 500~Hz for the three axes
(Figure \ref{f-mag_setup} a). Unlike the spacecraft-level test, we have several
flexibilities to investigate the coupling nature and possible mitigation: (1) we can
change the relative distance between the MTQ and the dewar arbitrarily to examine the
contribution of the magnetic coupling, and (2) we can test wrapping the harness between
the room-temperature electronics and the dewar with a magnetic shield material.

We had three configurations for the distance between the dewar center and EM MTQ: the
flight value (1135~mm) and 1.56 and 2.12 times of it. In each configuration, the MTQ was
driven in various duty ratios (20, 30, --70, and 80\%). Here, --70\% duty denotes 70\%
duty with the negative polarity ($-$35~V). The microcalorimeter bias setting was also
changed from nominal (1.6~V) to zero or a high (5.0~V) value. A part of this test was
repeated after wrapping the harness with the cobalt-based magnetic shield (2705M)
provided by Metglas$^{\textrm{\textregistered}}$, Inc\footnote{ See
\url{https://metglas.com/wp-content/uploads/2016/12/2705M-Technical-Bulletin.pdf} for
detail.}(Figure \ref{f-mag_setup} b).

\begin{figure}[htbp!]
 \centering
 \includegraphics[width=0.95\columnwidth,clip]{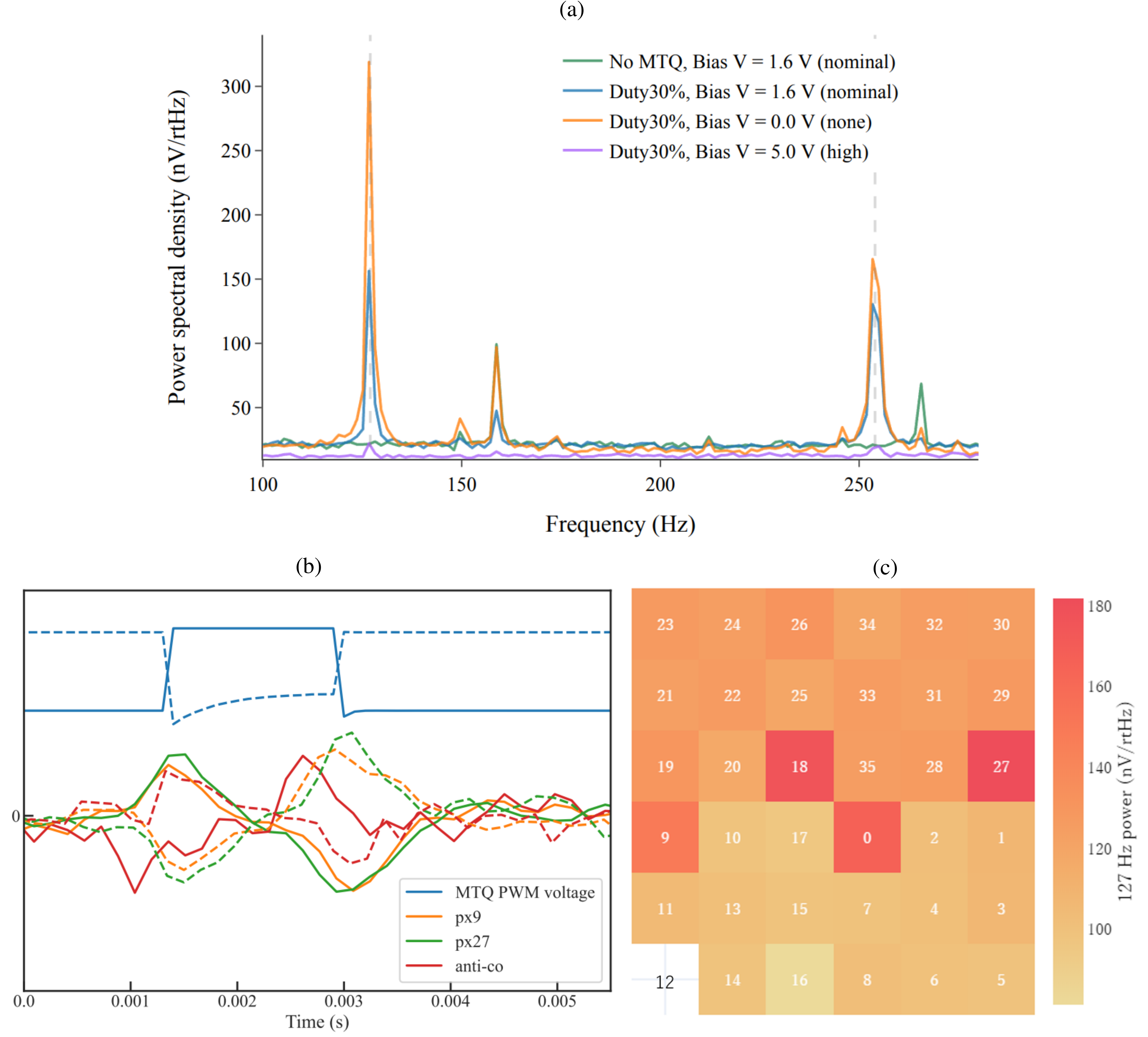}
 \caption{Results of the magnetic EMI instrument-level test: (a) Frequency-domain data of pixel 9 with varying detector bias (blue for
 nominal, violet for high, and orange for none) with a 30\% PWM duty of the MTQ. The PWM
 frequency (127~Hz) and its second harmonics are shown with dashed lines. Data of
 nominal bias without MTQ operation (green) is shown for comparison. The 150 and 156~Hz
 lines are, respectively, the third harmonic of the commercial AC and a cryocooler
 drive\cite{hasebe2022,imamura2022}. \label{f-mag_fd}
 (b) Time-domain data of the drive voltage (blue) and response of the
 microcalorimeter for pixel 9 (orange), 27 (green), and the anti-co (red) during the
 20\% (solid) and 80\% (dashed) duty drive. The anti-co signal is intrinsically opposite
 in polarity and shifted slightly in phase from that of the
 microcalorimeter. \label{f-mag_td}
 (c) Pixel dependence of the noise power at 127~Hz when the MTQ was operated
 with a 30\% PWM duty using the 8k noise spectra. Pixel 12 yielded no noise spectra by
 being interrupted by constant x-ray illumination by the $^{55}$Fe calibration
 source.\label{f-mag_pixel}}
 \label{f-mag_result1}
\end{figure}

Figure~\ref{f-mag_fd} shows the detector response in the frequency domain using the 8k
noise spectra with and without the MTQ with different detector biases. A strong
signature of the MTQ PWM frequency (127~Hz) and its harmonics was observed on both
calorimeter and anti-co channels when the MTQ was operated. For the anti-co, the
magnitude of the interference was independent of bias, whereas for the microcalorimeter
channels, the interference was stronger at zero bias than at nominal
bias. Figure~\ref{f-mag_td} shows the detector response in the time domain folded by the
PWM frequency using the bipolar power supply monitor reading for the input voltage (red)
and the microcalorimeter output for two selected pixels (blue and orange) as well as
anti-co's (green). The time between the input and output data set is shifted to match
the PWM edges to the microcalorimeter peaks and valleys as their relative time was not
calibrated. Figure~\ref{f-mag_pixel} shows the strength of the microcalorimeter noise
power at 127~Hz for all pixels in the array. The distribution is characterized by the
enhanced response in the pixels of a multiple of 9 (0, 9, 18, and 27) and an elevated
level in the upper half of the array than the lower half. These remarkable features were
consistently observed throughout the test. The measurement with
the magnetic shield wrapping the harness did not show significant difference beyond the
systematics.

\subsection{Spacecraft-level test}\label{s3-3}
We performed the magnetic EMI test during the spacecraft-level test on June 9, 2022. We
operated the three units of the MTQs separately for the $\pm$90 and $\pm$45\% duty ratio
and collected the detector noise spectra (a) to confirm the result obtained in the
instrument-level test (\S~\ref{s3-2}) and (b) to characterize the transfer function from
each unit of the MTQ to the microcalorimeter. We also operated the three MTQs with the
PWM duty ratio of ($x$, $y$, $z$)$=$($\pm$30, $\mp$30, $\pm$30)\% for an overnight
integration with x-rays and compared the result with another overnight result without
the MTQ to assess the degradation of the energy resolution by the presence of the strong
127~Hz and its harmonic lines in the noise spectra (Figure~\ref{f-mag_fd}). The duty
ratio was chosen to be the same among the three axes so that the time-domain peaks
(Figure~\ref{f-mag_td}) match for the worst case; the opposite polarity for $y$ is to
negate the opposite polarity of the MTQ-$y$ driver by design.

\begin{figure}[htbp!]
 \centering
 \includegraphics[width=0.9\columnwidth,clip]{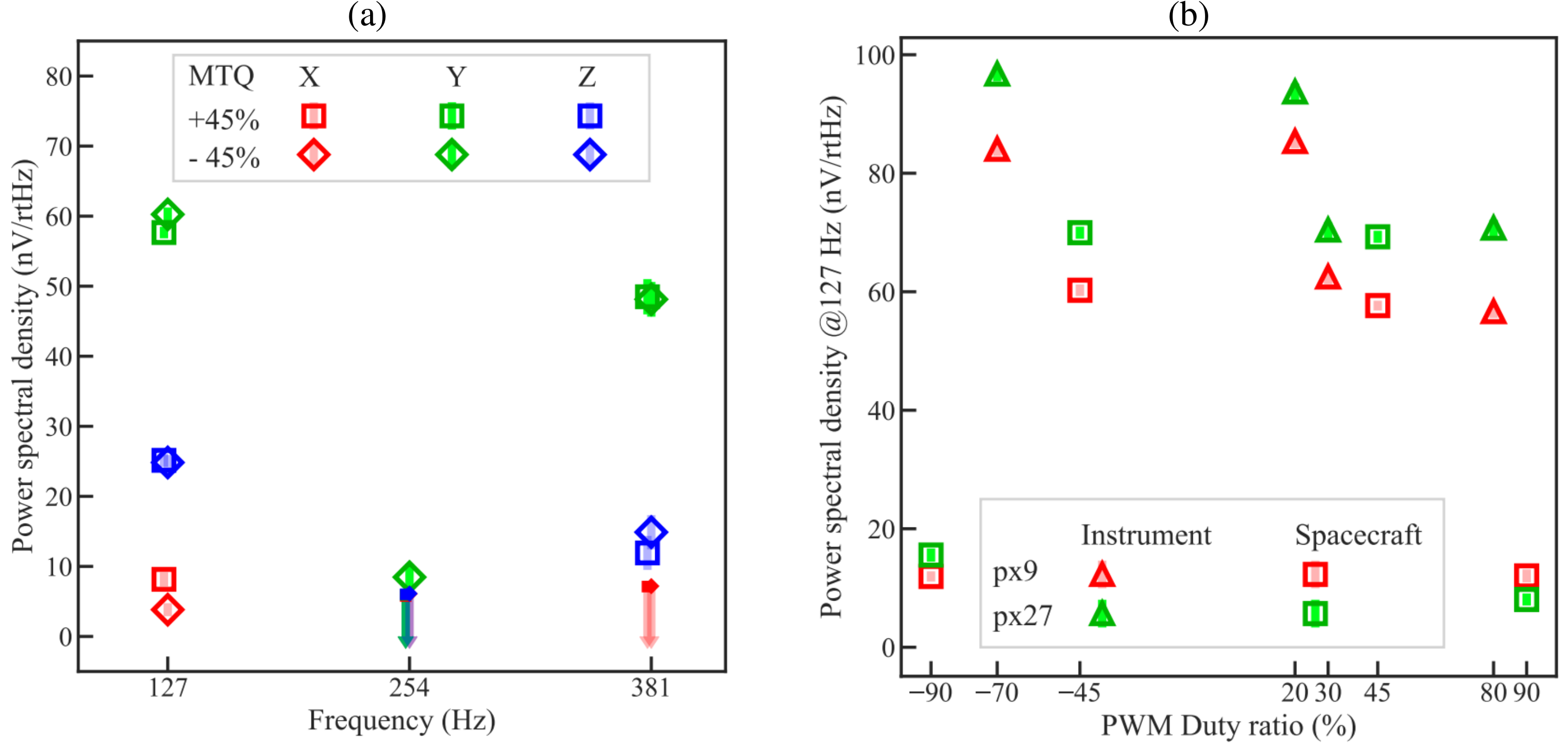}
 \caption{Results of the magnetic EMI spacecraft-level test: (a) Line power of px 9 at 127$n$~Hz ($n=$1, 2, 3) for the three MTQ units operated
 at $\pm 45$\% duty ratio. The underlying continuum was subtracted. The error
 bars are estimated from the continuum levels of the frequency neighbors. The upper
 limit of $3\sigma$ is given in case of no detection.\label{f-mag_unit}
 (b) Line power at 127~Hz for different duty ratio in the instrument- and
 spacecraft-level tests.\label{f-mag_duty}
 }
 \label{f-mag_result2}
\end{figure}

Figure~\ref{f-mag_unit} shows the microcalorimeter noise power at 127$n$~Hz ($n=$1, 2,
3) against the MTQ operation for each unit. The response was the strongest in the order
of $y$, $z$, and $x$. No significant difference was found between the opposite
polarities. Figure~\ref{f-mag_duty} shows the noise power at 127~Hz as a function of the
PWM duty ratio both in the instrument- and the spacecraft-level tests.

The MTQ duty ratio keeps changing in the orbit. We simulated its behavior, along with
the RW, during the spacecraft thermal-vacuum test in August 28--29, 2022 to assess their
impact in practice. The energy resolution of the microcalorimeter averaged over a
four-hour period with the MTQ and RW operation and another period without the MTQ and RW
operation was compared using calibration x-ray sources. No significant difference was
found and the risk of science impact is low. In case the increased coupling for some
reasons in the future, we have a backup option ready to notch the 127 and 254~Hz lines
in the onboard signal processing.

\subsection{Discussion}\label{s3-4}
Based on the results obtained in the instrument-level (\S~\ref{s3-2}) and
spacecraft-level (\S~\ref{s3-3}) tests and the simulation (\S~\ref{s3-1}), we speculate
how the MTQ couples with the microcalorimeter detector. We note that
the features observed in the \textit{Resolve} ground tests were very similar to 
those identified during SXS spacecraft-level tests, including 
the enhanced interference in specific pixels, the MTQ axis- and polarity-dependence 
of the detector response, the nature of the scaling with MTQ duty cycle, 
and the level of energy resolution degradation. The consistency between 
\textit{Resolve} and SXS test results lends support to the identification of 
the likely MTQ coupling mechanism outlined below.

\begin{figure}[htbp!]
 \centering
 \includegraphics[width=0.9\columnwidth,clip]{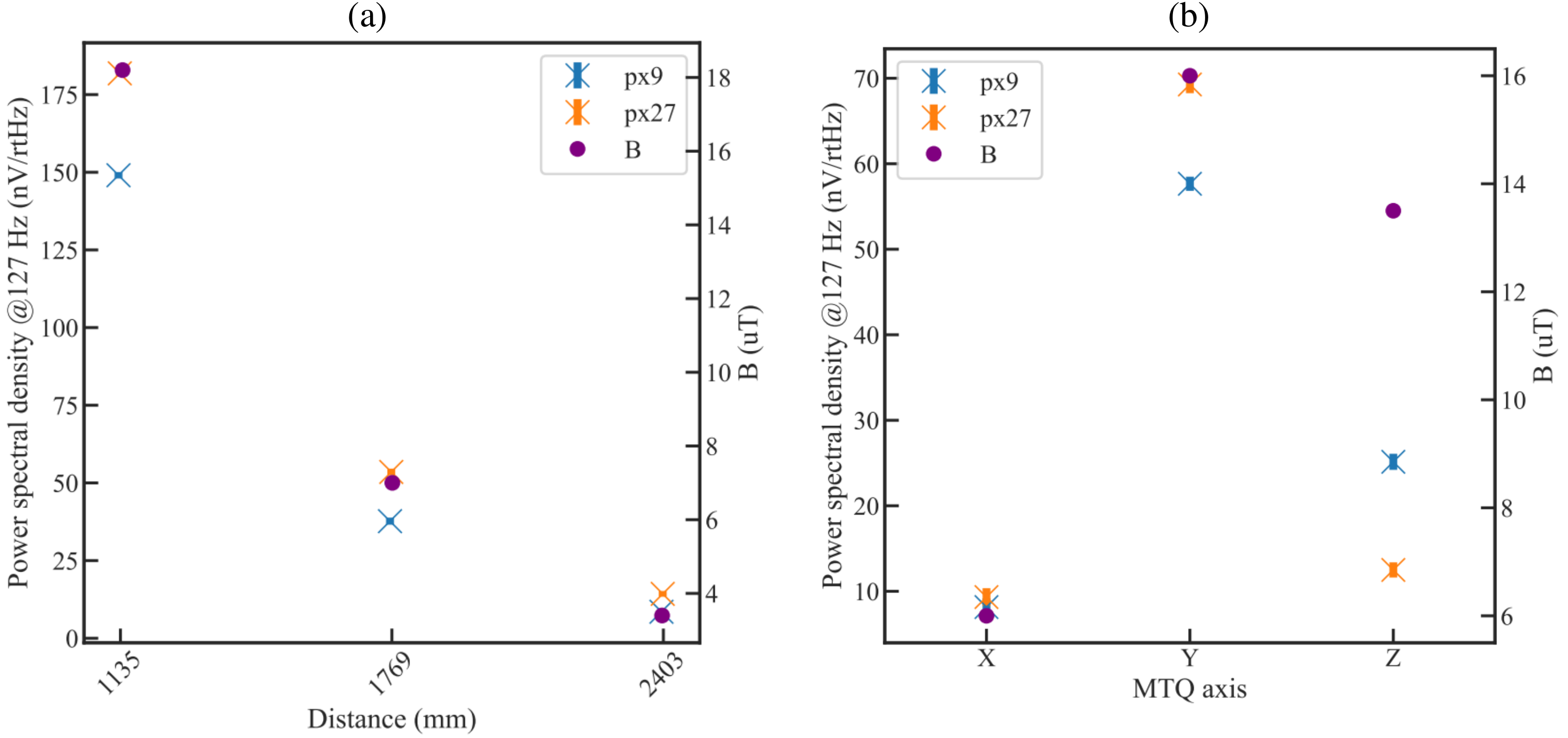}
 \caption{Comparison of the measurement and simulation: (a) Distance dependence of the 127~Hz power (pixels 9 and 27) and the
 simulated magnetic field strength $|\bm{B}|$ at the dewar center in the
 instrument-level test (\S~\ref{s3-2}) with the EM MTQ driven at a 30\%
 duty.\label{f-mag_dist}
 (b) Axis dependence of the 127~Hz power (pixels 9 and 27) and the simulated
 magnetic field strength $|\bm{B}|$ at the dewar center in the spacecraft-level test
 (\S~\ref{s3-3}) with the flight MTQs driven at a 45\% duty.\label{f-mag_axis}
 }
\end{figure}

First, the coupling is likely via the magnetic field. This is illustrated by the
distance dependence between the MTQ and the dewar in the instrument-level test
(Figure~\ref{f-mag_dist}). The dependence of the 127~Hz power in the detector noise
spectra decreases as the distance increases. Because other coupling mechanisms, such as
conductive coupling from the MTQ driver to the \textit{Resolve} room-temperature
electronics or low-frequency RF coupling from the MTQ driver, are not expected to
exhibit such dependence, we argue that the magnetic coupling is dominant. Indeed, the
observed distance dependence is quite similar to the simulated dependence of the field
strength $|\bm{B}|$ at the dewar center (Figure~\ref{f-mag_dist}). This explains why we
observed the strong axis dependence of the MTQs ($y$, $z$, and $x$ in the decreasing
order) in the spacecraft-level test (Figure~\ref{f-mag_unit}). It is mostly attributable
to the distance between each MTQ unit and the dewar (Figure~\ref{f-mag_axis}).

The magnetic coupling is also supported by the MTQ polarity dependence of the
response. The magnitude of response is the same between the two polarities
(Figure~\ref{f-mag_duty}) and the sign of the response is the opposite for opposite
polarity in the time domain (Figure \ref{f-mag_td}). Such behavior excludes the
possibility that the response is a function of the input power expected in conductive or
RF coupling.
 
We speculate that a particular part of the instrument is sensitive to a particular
direction of the magnetic field. We injected local magnetic field using a portable
solenoid driven with a 127~Hz sine wave and found that the harness between the
room-temperature electronics (ADRC, XBOX) and the dewar and their ends are particularly
sensitive. There might be other susceptible places, in particular inside the dewar, which was not
accessible with the portable solenoid.

If the nearly DC magnetic field is the path for the coupling, why do we observe an AC
response in the microcalorimeter? This can be studied by using the time-domain data in
Figure~\ref{f-mag_td}. The spacing between rising and falling edges of the input voltage
$V_{\mathrm{input}}(t)$ coincides with that between the peak and the valley of the
microcalorimeter output in all tested duty cycles. The magnetic field is smoothed in
time for its large inductance with $B(t) \propto \int V_{\mathrm{input}}(t) dt$, but the
induced voltage is $V_{\mathrm{ind}}(t) \propto dB(t)/dt \propto
V_{\mathrm{input}}(t)$. If this further couples capacitively, the noise voltage would be
$V_{\mathrm{noise}}(t) \propto dV_{\mathrm{ind}}(t)/dt \propto
dV_{\mathrm{input}}(t)/dt$. This explains the edge--peak relation in the time-domain
data. The microcalorimeter response is smaller for the 90\% duty than the 45\% duty in
the spacecraft-level test (Figure~\ref{f-mag_duty}). This is probably because the rising
and falling edges are close in time for the 90\% duty so they cancel to some extent when
arithmetically added.

The likely place of the capacitive coupling can be investigated using the pixel
dependence in Figure~\ref{f-mag_pixel}. The pixels with a multiple of 9 (pixels 0, 9,
18, and 27) are particularly susceptible to the magnetic EMI. This is related to the
readout wire layout from the microcalorimeter\cite{Chiao2018}. Figure~\ref{f-spacecraft}
(d) shows the close-up view, in which two bundles of wires (18 pairs of signal and
return for each pixel) come out of the detector array. One bundle reads the upper 
half of the array and the other the lower half. This part with a high impedance before
the JFET impedance conversion is particularly sensitive to external radiative noise,
hence a likely site of the capacitive coupling. The signal and return wires are 
aligned alternately, and all signal wires but the outermost (pixels 0, 9, 18, 27) are 
in between grounded return wires. This could explain the peculiar pixel dependence 
of magnetic coupling. 

We have ample evidence that the magnetic interference is not acting through heating, but
is purely electrical. The first indication is that the pickup is seen on the anti-co
(Figure~\ref{f-mag_td}), which is a non-thermal device. Furthermore, the signal in the
time domain is bipolar (Figure~\ref{f-mag_td}). Finally, the detector bias dependence
(Figure~\ref{f-mag_fd}) is not consistent with heating.  With no bias, the
microcalorimeter would not work as a thermal detector. Still, the strong MTQ noise lines
were observed, illustrating the electrical nature of the input. This is in contrast to
the thermal nature of the input found in the low-frequency noise by the cryocooler
micro-vibration interference of the \textit{Resolve}
instrument\cite{hasebe2022,imamura2022}. In fact, the dependence of the magnetic
coupling on bias for a particular pixel appears to scale with the expected temperature-
and frequency-dependent impedance of the thermistor.


\section{RF EMI}\label{s4}
\subsection{Simulation}\label{s4-1}
For RF EMI, we also start with the simulation, which requires massive computational
resources unlike the low-frequency magnetic EMI (\S~\ref{s3-1}). RF simulation is
typically performed with a mesh size $\sim$1/20 of the wavelength (i.e., 2~mm for the
X-band) with a model detailed to the scale. This is far smaller than the spacecraft
size. Therefore, in spacecraft RF simulations, simplified models with hybrid solvers are
often used. In this study, however, we use the detailed CAD model with a single solver
for simulating the entire spacecraft.

For the solver, we use the finite difference time domain (FDTD)
method\cite{sullivan2013}. This solves a discretized Maxwell's equations on the Yee's
lattice\cite{yee1966}. We adopt the \texttt{Poynting for Microwave}
software\footnote{See
\url{https://www.fujitsu.com/jp/solutions/business-technology/tc/sol/poynting/} for
detail.} by Fujitsu. For the computer, we use Fugaku\cite{Sato2020}, a high-performance
computing facility in RIKEN, which is also manufactured by Fujitsu. The most detailed
CAD model for the entire ASTRO-H spacecraft was used with some modifications in
accordance with the design changes for the XRISM. The materials are replaced with the
perfect electric conductor. The perfect matched layer was set as the boundary condition
for the simulation box (Figure~\ref{f-simulation} b).

\begin{figure}[htbp!]
 \centering
 \includegraphics[width=0.9\columnwidth,clip]{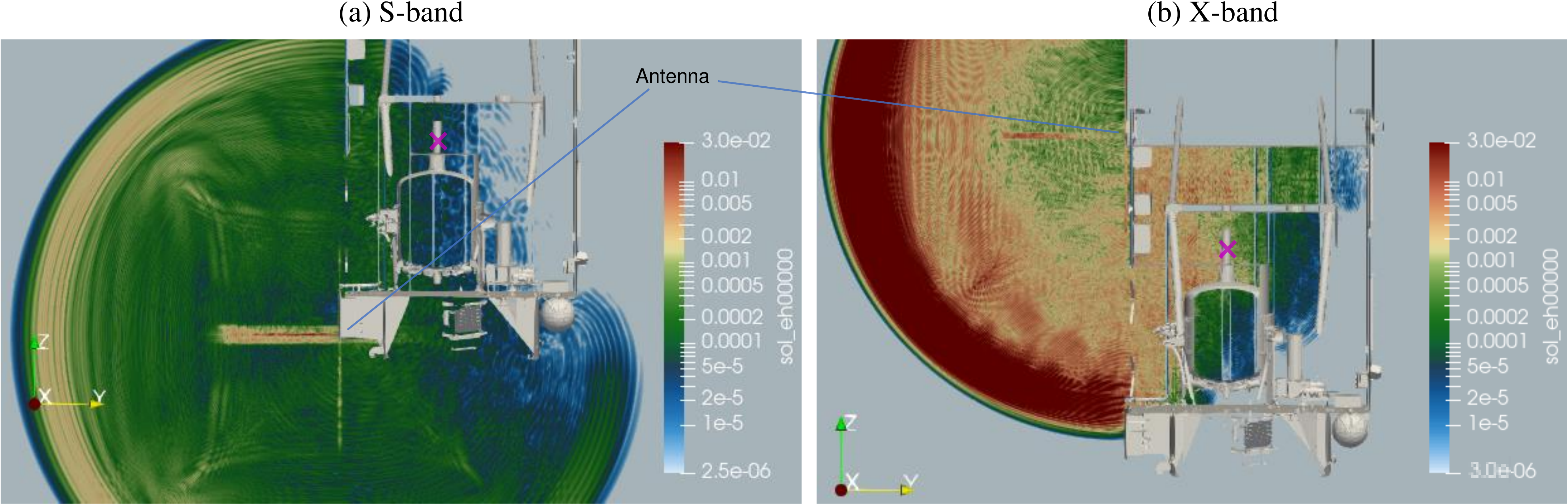}
 \caption{Results of the RF simulation for the (a) S-band and (b) X-band antenna on SP7.
 The field strength $|\bf{E}|$ is given in the unit of V~m$^{-1}$ on the plane vertical
 to the $x$-axis including the detector (Figure~\ref{f-simulation} b). The interface is
 marked with the magenta cross. The horizontal structure from the antennas is an
 artifact of the simulation. \texttt{Paraview}\cite{ahrens2005} is used for
 visualization.}
 \label{f-rfsim2}
\end{figure}

We injected the maximum operational power used for transmission (31.6/36.3~dBm
respectively for the S/X-band antenna outside of SP7; the closer to the dewar of the
two for the S-band antenna). The simulation was run for 25/23~$\mu$s in time and a
total mesh of $1.9 \times 10^{11}$ in space using 1024 nodes of Fugaku for 3 hours. A
snapshot at 11.1/10.2~ns is shown in Figure~\ref{f-rfsim2}. The calculated field
strength at the interface above the GV is $1.5 \times 10^{-4}$/$5.6 \times 10^{-3}$
V~m$^{-1}$, which corresponds to --106/--63 dBm, respectively for the S/X-band.

\subsection{Instrument-level test}\label{s4-2}
\begin{figure}[htbp!]
 \centering
 \includegraphics[width=0.95\columnwidth,clip]{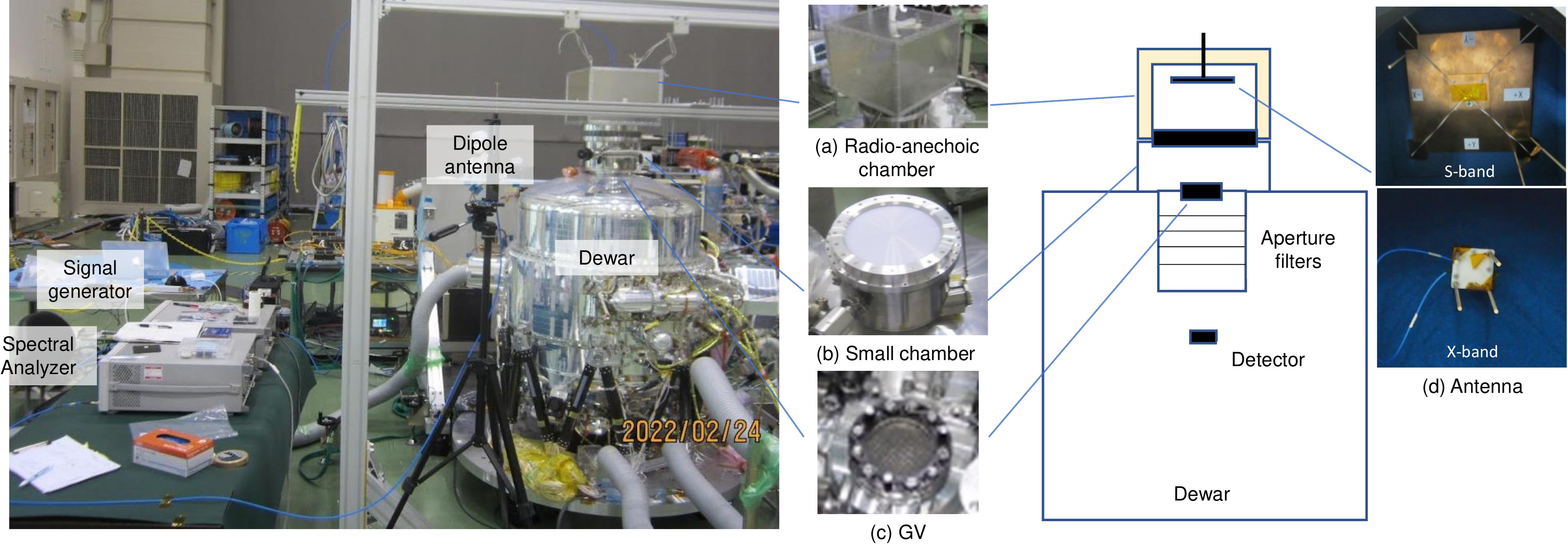} 
\caption{Setup for the RF EMI instrument-level test. (Left) photo of the
 setup. (Right) configuration and close-up views of some major apparatuses.}
 \label{f-rf_setup}
\end{figure}

We performed the RF EMI test during the instrument-level test using the flight-model
hardware on February 24 and 28, 2022. The test was designed to measure the detector
response against the RF injection at the S and X-band from above the dewar by opening
the GV. Figure~\ref{f-rf_setup} shows the experimental setup. We avoided moving the
entire instrument to an EMC test room in a different building to avoid various risks,
thus the test was performed in a clean room continued from other tests. Special
apparatuses described below for this test are not mechanically compatible with the
spacecraft structure, so the instrument-level is the highest level of integration to
perform this test.

In the setup, we need to meet two requirements: one is to keep the dewar vacuum and the
other is to comply with the radio act of the government. For the former, we used a small
vacuum chamber to cover the GV and kept the dewar leak-tight even when the GV was opened
(Figure~\ref{f-rf_setup} b\cite{ishisaki2022}). The GV can be opened and closed
repeatedly using a handle on the ground. The top part of the chamber was replaced with
an RF transmissive window made of high-density polyethylene. For the latter, a
radio-anechoic chamber made of Al with the radio absorber interior was placed on top of
the chamber in the air (Figure~\ref{f-rf_setup} a).  The S- or X-band antenna was placed
inside (Figure~\ref{f-rf_setup} d) to emit power provided by a signal generator toward
the microcalorimeter. The S-band antenna was made and characterized in-house and the
X-band antenna was borrowed from the OMOTENASHI project\cite{Hashimoto2019} using a
close frequency. A dipole antenna was placed 3~m away from the radio anechoic chamber to
monitor the RF leakage using a spectral analyzer. As the RF injection path is only
through the opened GV, the small radio-anechoic chamber suffices for the test purpose.

\begin{figure}[htbp!]
 \centering
 \includegraphics[width=0.98\columnwidth,clip]{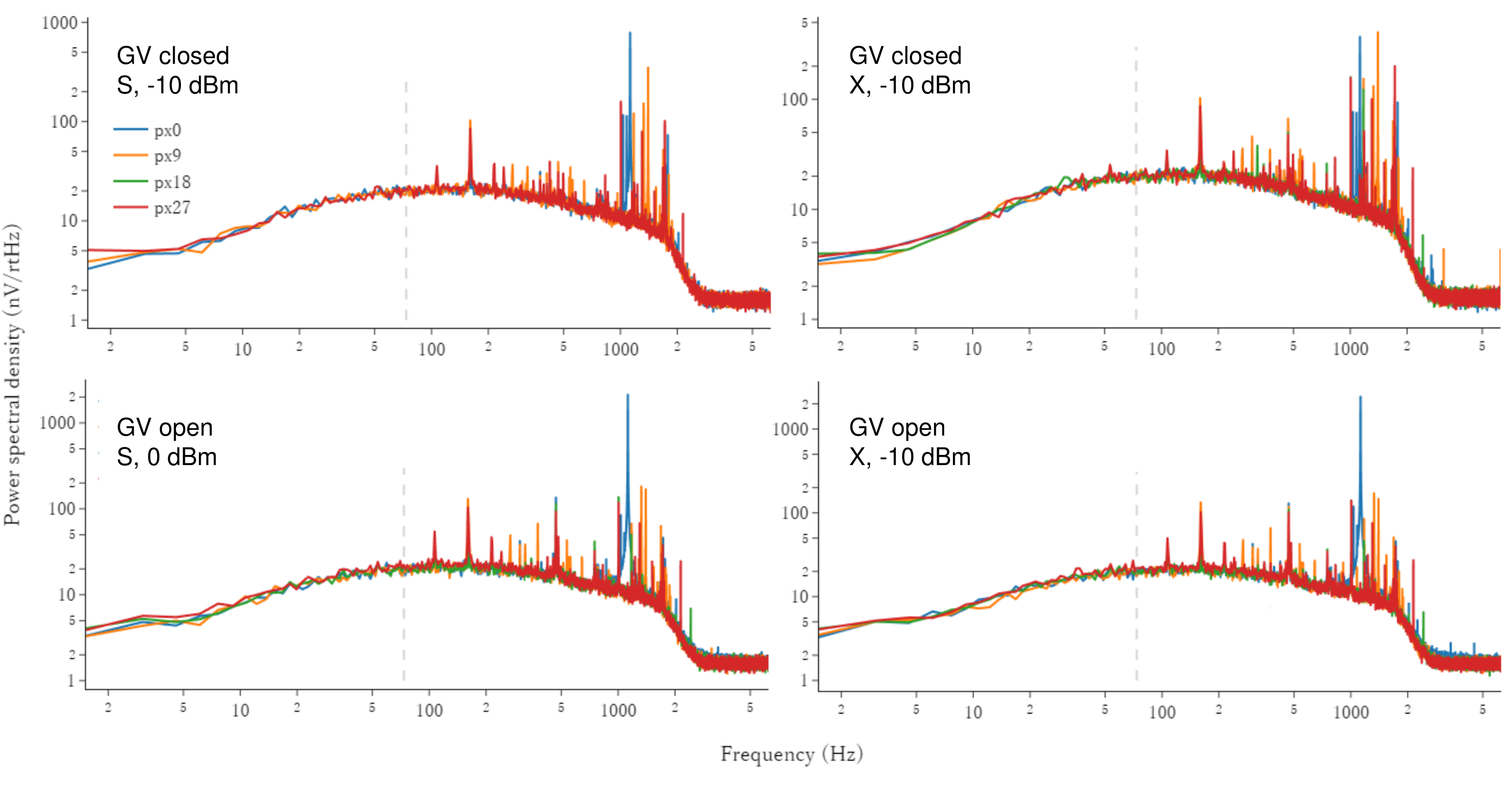}
 \caption{Results of the RF EMI instrument-level test. 8k noise spectrum of pixel
 0, 9, 18, and 27 against the strongest RF injection. The dashed line shows 73.5~Hz.}
 \label{f-rf_ns}
\end{figure}

The injection level was increased from --120~dBm to 0~dBm by a 20~dB increment for the S
and X-band in both the GV closed and open configurations. When the monitor was about to
exceed the legal limit, the strongest injection was replaced with --10~dBm. The carrier
frequency was amplitude-modulated at 73.5~Hz so that the modulated power is visible in
the detector bandpass. At each injection, we obtained the 8k noise spectra of the
microcalorimeter and measured the power at 73.5~Hz. Figure~\ref{f-rf_ns} shows the 8k
noise spectra of some selected pixels in the strongest injection case of each
configuration. No significant excess noise at the modulation frequency was observed in
any of the measurements. The upper limit was 15--20~nV/$\sqrt{\mathrm{Hz}}$, which is
negligible for any degradation of the detector performance.
No interference was detected on the anti-co signal either.

\subsection{Spacecraft-level test}\label{s4-3}
\begin{figure}[htbp!]
 \centering
 \includegraphics[width=0.4\columnwidth,clip]{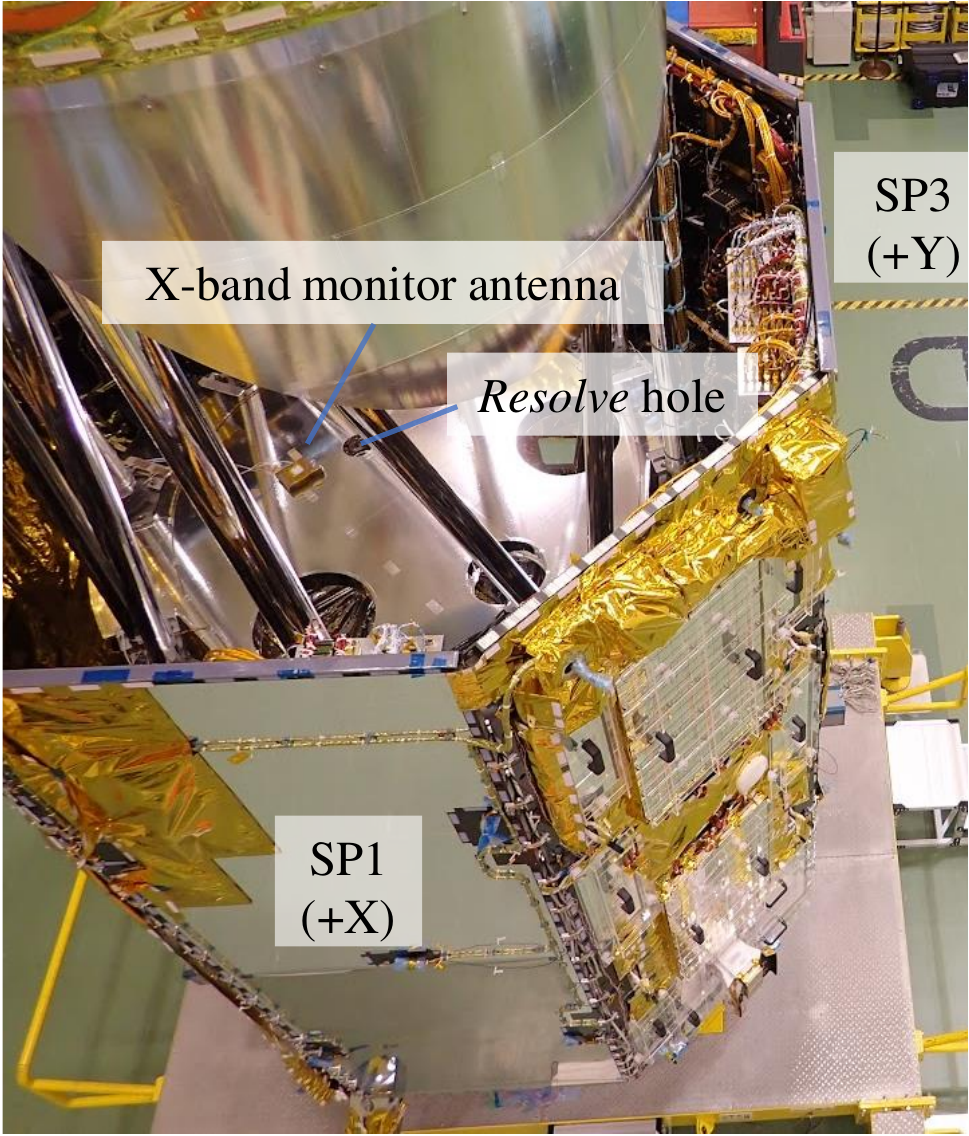}
 \caption{Setup for the RF EMI spacecraft-level test. An X-band antenna was installed on
 top of the lower plate close to the hole in the \textit{Resolve} optical path. The
 photo was taken from the top of SP1 and SP2 in the downward-looking direction before
 the integration was completed.}
 \label{f-rf_setup2}
\end{figure}

We cannot open the GV in the spacecraft configuration on the ground, so no end-to-end
assessment is possible when the spacecraft RF system is operating. We placed the X-band
antenna, which was used as a transmitter in the instrument-level test (\S~\ref{s4-3})
and used it as a receiver in the spacecraft-level test, at a place close to the dewar
entrance (Figure~\ref{f-rf_setup2}) and monitored the level of the field inside the
spacecraft during the spacecraft-level tests. When the RF systems were operated in
various modes for the air-link communication testing in June, 2022, the maximum measured
level was about --80~dBm, which validates the assumed input level in the
instrument-level test (\S~\ref{s4-2}) based on the simulation (\S~\ref{s4-1}).

\subsection{Discussion}\label{s4-4}
By combining the simulation and tests, we found that the microcalorimeter in
\textit{Resolve} was found immune to the RF input from the open hole in the top of the
dewar when the GV is opened. The RF input of 1~mW, which corresponds to the maximum
injection power at the instrument-level test (\S~\ref{s4-2}), is larger by many orders
than the x-ray input from the same direction during observations of
$\mathcal{O}(1~\mathrm{fW})$ for a milli-Crab source. The microcalorimeter is protected
against the input in electromagnetic forms of longer wavelengths than x-rays by the 
multiple layers of thin filters made of aluminized polyimide in the x-ray aperture\cite{Kilbourne2018}.
For RF input, the Al in the filters of a 50--100~nm thickness
reflects back most of the incoming power at its surface by the impedance mismatch with
the vacuum, allowing only a small fraction (less than --50~dB) to penetrate. We
speculate that multiple layers of such filters work effectively to reject RF
noise input.

\section{Summary}\label{s5}
We presented the results of the ground testing and simulation of the EMI to the x-ray
microcalorimeter in the \textit{Resolve} instrument onboard the XRISM. We discussed the
magnetic EMI caused by the MTQs in the spacecraft attitude control system and the RF EMI
caused by the spacecraft communication system. For the magnetic EMI, we observed a
strong response in the microcalorimeter and anti-co time- and frequency-domain data. We
speculated its coupling mechanisms based on the test and simulation results. There is no
evidence that the resultant degradation is beyond the current allocation of noise
budget.  For the RF EMI, we injected RF signal up to 0/--10 dBm for the S/X-band but did
not observe any response in the microcalorimeter and anti-co data in the
instrument-level test. Because no end-to-end assessment is possible for the RF EMI on
the ground, we conducted a full RF simulation using a detailed spacecraft CAD model and
a single solver with Fugaku to complement the limitation. We found that the expected
levels of the field diffracted from the spacecraft antennas are much smaller (--106/--63
dBm for S/X-band) than the level that \textit{Resolve} was tested to be immune.

\subsection* {Acknowledgments}
This work is made possible with significant contributions of all the XRISM
\textit{Resolve} team members, and the SHI and NEC engineers, which we greatly
appreciate. Kenji Fukunabe, Yoshiaki Mitsutake, Atsushi Tomiki, Yutaro Sekimoto, Hayato
Takakura at ISAS helped us to prepare for the instrument-level tests. This work used
computational resources of the supercomputer Fugaku provided by RIKEN. This paper is a derivation of the SPIE proceeding\cite{kurihara2022}.


\bibliography{report}   

\begin{thebibliography}{10}

\bibitem{tauber2010}
J.~A. Tauber {\em et~al.}, ``Planck pre-launch status: {{The Planck}}
  mission,'' {\em antike und abendland} {\bf 520}, A1  (2010).

\bibitem{lamarre2010}
J.~M. Lamarre {\em et~al.}, ``Planck pre-launch status: {{The HFI}} instrument,
  from specification to actual performance,'' {\em antike und abendland} {\bf
  520}, A9  (2010).

\bibitem{gualtieri2018}
R.~Gualtieri {\em et~al.}, ``{{SPIDER}}: {{CMB}} polarimetry from the edge of
  space,'' {\em Journal of Low Temperature Physics} {\bf 193}, 1112--1121
  (2018).

\bibitem{takahashi2016}
T.~Takahashi {\em et~al.}, ``The {{ASTRO-H}} (hitomi) x-ray astronomy
  satellite,'' in {\em Space Telescopes and Instrumentation 2016:
  {{Ultraviolet}} to Gamma Ray},  J.-W.~A. {den Herder}, T.~Takahashi, and
  M.~Bautz, Eds.,  {\bf 9905}, 99050U, {SPIE}  (2016).

\bibitem{kelley2016}
R.~L. Kelley {\em et~al.}, ``The {{Astro-H}} high resolution soft x-ray
  spectrometer,'' in {\em Space Telescopes and Instrumentation 2016:
  {{Ultraviolet}} to Gamma Ray},  J.-W.~A. {den Herder}, T.~Takahashi, and
  M.~Bautz, Eds.,  {\bf 9905}, 99050V  (2016).

\bibitem{mitsuda2014}
K.~Mitsuda {\em et~al.}, ``Soft x-ray spectrometer ({{SXS}}): The
  high-resolution cryogenic spectrometer onboard {{ASTRO-H}},'' in {\em Space
  Telescopes and Instrumentation 2014: {{Ultraviolet}} to Gamma Ray},
  T.~Takahashi, J.-W.~A. {den Herder}, and M.~Bautz, Eds.,  {\bf 9144}, 91442A
  (2014).

\bibitem{ishisaki2022}
Y.~Ishisaki, R.~L. Kelley, H.~Awaki, {\em et~al.}, ``Status of resolve
  instrument onboard x-ray imaging and spectroscopy mission ({{XRISM}}),'' in
  {\em Space {{Telescopes}} and {{Instrumentation}} 2022: {{Ultraviolet}} to
  {{Gamma Ray}}},   {\bf 12181}, 409--430, {SPIE}  (2022).

\bibitem{tashiro2020}
M.~S. Tashiro {\em et~al.}, ``Status of x-ray imaging and spectroscopy mission
  ({{XRISM}}),'' in {\em Space Telescopes and Instrumentation 2020:
  {{Ultraviolet}} to Gamma Ray},  J.-W.~A. {den Herder}, K.~Nakazawa, and
  S.~Nikzad, Eds.,  {\bf 11444}, 176, {SPIE}  (2020).

\bibitem{eckart2018}
M.~E. Eckart, J.~S. Adams, K.~R. Boyce, {\em et~al.}, ``Ground calibration of
  the {{Astro-H}} ({{Hitomi}}) soft x-ray spectrometer,'' in {\em Journal of
  {{Astronomical Telescopes}}, {{Instruments}}, and {{Systems}}},
  T.~Takahashi, J.-W.~A. {den Herder}, and M.~Bautz, Eds.,  {\bf 4}, 1, {SPIE}
  (2018).

\bibitem{porter2018}
F.~S. Porter, K.~R. Boyce, M.~P. Chiao, {\em et~al.}, ``In-flight performance
  of the soft x-ray spectrometer detector system on {{Astro-H}},'' {\em Journal
  of Astronomical Telescopes, Instruments, and Systems} {\bf 4}, 1  (2018).

\bibitem{leutenegger2018}
M.~A. Leutenegger {\em et~al.}, ``In-flight verification of the calibration and
  performance of the {{ASTRO-H}} ({{Hitomi}}) {{Soft X-ray Spectrometer}},''
  {\em Journal of Astronomical Telescopes, Instruments, and Systems} {\bf 4},
  021407  (2018).

\bibitem{kilbourne2018a}
C.~A. Kilbourne {\em et~al.}, ``Design, implementation, and performance of the
  {{Astro-H SXS}} calorimeter array and anticoincidence detector,'' {\em
  Journal of astronomical telescopes, instruments, and systems} {\bf 4}(1),
  011214  (2018).

\bibitem{Chiao2018}
M.~P. Chiao {\em et~al.}, ``System design and implementation of the detector
  assembly of the {{Astro-H}} soft x-ray spectrometer,'' {\em Journal of
  Astronomical Telescopes, Instruments, and Systems} {\bf 4}(2), 1--16  (2018).

\bibitem{ishisaki2018}
Y.~Ishisaki {\em et~al.}, ``Resolve instrument on {{X-ray}} astronomy recovery
  mission ({{XARM}}),'' {\em Journal of Low Temperature Physics} {\bf 193},
  991--995  (2018).

\bibitem{fujimoto2017}
R.~Fujimoto {\em et~al.}, ``Performance of the helium dewar and the cryocoolers
  of the {{Hitomi}} soft x-ray spectrometer,'' {\em JATIS} {\bf 4}, 1  (2017).

\bibitem{yoshida2018}
S.~Yoshida, ``Cooling system for the soft {{X-ray}} spectrometer onboard the
  {{ASTRO-H}},'' {\em TEION KOGAKU (Journal of Cryogenics and Superconductivity
  Society of Japan)} {\bf 53}, 349--354  (2018).

\bibitem{shirron2016}
P.~J. Shirron, M.~O. Kimball, B.~L. James, {\em et~al.}, ``Design and on-orbit
  operation of the adiabatic demagnetization refrigerator on the {{Hitomi Soft
  X-ray Spectrometer}} instrument,'' in {\em Space {{Telescopes}} and
  {{Instrumentation}} 2016: {{Ultraviolet}} to {{Gamma Ray}}},  J.-W.~A. {den
  Herder}, T.~Takahashi, and M.~Bautz, Eds.,  {\bf 9905}, 99053O,
  {International Society for Optics and Photonics}  (2016).

\bibitem{sato2012}
Y.~Sato {\em et~al.}, ``Development of mechanical cryocoolers for the cooling
  system of the {{Soft X-ray}} spectrometer onboard {{Astro-H}},'' {\em
  Cryogenics} {\bf 52}, 158--164  (2012).

\bibitem{sneiderman2018}
G.~A. Sneiderman {\em et~al.}, ``Cryogen-free operation of the {{Soft X-ray
  Spectrometer}} instrument,'' {\em JATIS} {\bf 4}, 1  (2018).

\bibitem{kanao2017}
K.~Kanao {\em et~al.}, ``Cryogen free cooling of {{ASTRO-H SXS}} helium dewar
  from 300 {{K}} to 4 {{K}},'' {\em Cryogenics} {\bf 88}, 143--146  (2017).

\bibitem{Midooka2021}
T.~Midooka {\em et~al.}, ``X-ray transmission calibration of the gate valve for
  the x-ray astronomy satellite {{XRISM}},'' {\em Journal of Astronomical
  Telescopes, Instruments, and Systems} {\bf 7}(2), 028005  (2021).

\bibitem{hasebe2022}
T.~Hasebe, R.~Imamura, M.~Tsujimoto, {\em et~al.}, ``Ground test results of the
  micro vibration interference for the x-ray microcalorimeter onboard
  {{XRISM}},'' in {\em Space {{Telescopes}} and {{Instrumentation}} 2022:
  {{Ultraviolet}} to {{Gamma Ray}}},   {\bf 12181}, 1527--1539, {SPIE}  (2022).

\bibitem{imamura2022}
R.~Imamura, M.~Tsujimoto, H.~Awaki, {\em et~al.}, ``Results of accelerometer
  monitor in the ground testing of {{Resolve}} x-ray microcalorimeter
  instrument onboard {{XRISM}},'' in {\em X-{{Ray}}, {{Optical}}, and
  {{Infrared Detectors}} for {{Astronomy X}}},   {\bf 12191}, 763--770, {SPIE}
  (2022).

\bibitem{sullivan2013}
D.~M. Sullivan, {\em Electromagnetic Simulation Using the {{FDTD}} Method},
  {John Wiley \& Sons}  (2013).

\bibitem{yee1966}
K.~Yee, ``Numerical solution of initial boundary value problems involving
  {{Maxwell}}'s equations in isotropic media,'' {\em IEEE Transactions on
  antennas and propagation} {\bf 14}(3), 302--307  (1966).

\bibitem{Sato2020}
M.~Sato {\em et~al.}, ``Co-design for a64fx manycore processor and''
  fugaku'','' in {\em {{SC20}}: {{International}} Conference for High
  Performance Computing, Networking, Storage and Analysis},  1--15, {IEEE}
  (2020).

\bibitem{ahrens2005}
J.~Ahrens, B.~Geveci, and C.~Law, ``{{ParaView}}: {{An End-User Tool}} for
  {{Large-Data Visualization}},'' in {\em Visualization {{Handbook}}},  C.~D.
  Hansen and C.~R. Johnson, Eds., 717--731, {Butterworth-Heinemann},
  {Burlington}  (2005).

\bibitem{Hashimoto2019}
T.~Hashimoto {\em et~al.}, ``Nano semihard moon lander: {{OMOTENASHI}},'' {\em
  IEEE Aerospace and Electronic Systems Magazine} {\bf 34}(9), 20--30  (2019).

\bibitem{Kilbourne2018}
C.~A. Kilbourne {\em et~al.}, ``Design, implementation, and performance of the
  {{Astro-H}} soft x-ray spectrometer aperture assembly and blocking filters,''
  {\em Journal of Astronomical Telescopes, Instruments, and Systems} {\bf
  4}(1), 011215  (2018).

\bibitem{kurihara2022}
M.~Kurihara, M.~Tsujimoto, M.~Eckart, {\em et~al.}, ``Ground test results of
  the electromagnetic interference in the x-ray microcalorimeter onboard
  {{XRISM}},'' in {\em Space {{Telescopes}} and {{Instrumentation}} 2022:
  {{Ultraviolet}} to {{Gamma Ray}}},   {\bf 12181}, 1445--1458, {SPIE}  (2022).

\end{thebibliography}
\bibliographystyle{spiejour}   


\vspace{2ex}\noindent\textbf{Miki Kurihara} is a graduate student of the University of
Tokyo (astronomy major) working for XRISM at JAXA/ISAS. He received his BS in physics
from Chiba University in 2021. He is a student member of SPIE.

\listoffigures

\end{spacing}
\end{document}